\newcommand{\sigmath}{\sigma_\mathrm{T}}

\newcommand{\kte}{kT_\mathrm{e}}
\newcommand{\me}{m_\mathrm{e}}
\newcommand{\mcc}{\me c^{2}}
\newcommand{\taus}{\tau_\mathrm{s}}
\newcommand{\taut}{\tau_\mathrm{T}}

\documentclass{aa} 

\usepackage{aas_macros,graphics}
\begin{document}

   \thesaurus{02     
              (02.01.2;  
               02.18.5;  
               03.13.4;  
               08.02.3;  
               11.19.1;  
               13.25.3)} 

   \title{Temporal properties of flares in accretion disk  coronae}

   \subtitle{}
 
   \author{J. Malzac \and
           E. Jourdain
          }

   \offprints{J. Malzac}

   \institute{Centre d'Etude Spatiale des Rayonnements (CNRS/UPS)
              9, Av du Colonel Roche, 31028, Toulouse Cedex 4, France\\
              email:Julien.Malzac@cesr.fr
             }

   \date{Received ??, 1999; accepted ??, 1999}
  \titlerunning{Temporal properties}
   \maketitle
 
   \begin{abstract}

Using a non-linear Monte-Carlo code we investigate the radiative
 response of an accretion disk corona system to static homogeneous 
flares. We model a flare by a rapid (comparable to the light crossing time) 
energy dissipation 
in the corona or the disk.

If the flares originate from the disk,
 the coronal response to the soft photon shots produces a strongly
 non-linear Comptonised 
radiation output, with  complex correlation/anti-correlations between 
energy bands. This behavior strongly differs from those found with usual
 linear calculations. Thus any model for the  rapid aperiodic 
variability of X-ray binaries 
invoking a varying soft photon input as a source
for the hard X-ray variability has to take into account the 
coronal temperature response.  

On the other hand, if the flare is due to a violent heating of the corona,      when the perturbation time scale is of the order of a few 
corona light crossing times,
 the shot spectrum evolves from hard to soft. This general trend is
 independent of 
the shot profile and  geometry. We show that for short dissipation time,
 the time averaged
spectra are generally harder than in steady state situation. In addition,
 annihilation line and high energy tails can be produced without need 
for non-thermal processes.

   \end{abstract}


\section{Introduction}
The hard X/$\gamma$-ray spectra of  galactic black hole candidates (GBHC)
 in their low hard state as well as Seyfert galaxies
  can be generally represented  as a sum of a hard
  power-law  continuum  with a cutoff  at a few  hundred  keV and a  Compton
  reflection  bump (with a Fe  K$\alpha$  line at $\sim 6.4$ keV)  produced
  when high energy  photons  interact with cold  material (Zdziarski et al. 1996\nocite{1996A&AS..120C.553Z}).  The presence of
  the Compton  reflection  component  (Guilbert \& Rees 1988\nocite{1988MNRAS.233..475G}; George \& Fabian 1991\nocite{1991MNRAS.249..352G}) implies that
  cold  material  could be
  present in the direct vicinity of the X/$\gamma$-ray producing region.  A
  soft excess  present  below $\sim 1$ keV is usually  associated  with the
  thermal  emission from the cold accretion  disk and can be powered by the
  viscous  dissipation in the disk itself as well as by  reprocessing  hard
  photons.  These  three   components  are  generally   interpreted  in  the
  framework of accretion disk corona  models.  These models assume that the
  soft thermal  radiation is emitted by the disk and then  Comptonised in a
  very hot plasma,  the  ``corona''. The reflection features arise naturally from the disk illumination. The nature and  geometry  of the corona are
  unclear.  Several   geometries  have  been  proposed: slab sandwich-like corona (Haardt \& Maraschi, 1991, 1993)\nocite{1993ApJ...413..507H}\nocite{1991ApJ...380L..51H},  localized   active
  regions on the disk surface powered by magnetic reconnections 
(Liang et al. 1977; \nocite{1977ApJ...218..247L}Galeev et al. 1979; \nocite{1979ApJ...229..318G} Haardt et al. 1994\nocite{1994ApJ...432L..95H})  or a hot accretion
  disk  in  the  center  surrounded  by a  cold  standard disk (Shapiro et al. 1976 \nocite{1976ApJ...204..187S}; Ichimaru 1977\nocite{1977ApJ...214..840I}; Narayan \& Yi 1994\nocite{1994ApJ...428L..13N}).
  Unfortunately, spectroscopy alone does not allow one to firmly probe either
  the geometry or the way the corona is powered.

 X-ray  variability  studies  seem to be a key for  the  understanding  of
  accretion  processes around compact  objects.  They should at least bring
  indications  on the nature and  structure of the corona. The short term variability of black hole X-ray binaries is now well known thanks to an 
  impressive amount of observational  data accumulated  since many years by
  several space  experiments (Cui 1999)\nocite{1999hepa.conf...97C}.  There is however  no accepted  model that accounts for
  most  of  the  observational   data.  The  inner  disk  dynamic  predicts
  important  variability at kHz frequencies while a stronger variability is
  observed around 1~Hz.  The power density  spectrum is roughly a power-law
  of index between -1.0 and -2.0.  Another  important  feature of the  temporal
  behavior is the time lags between hard and soft  photons  that depend on
  Fourier frequency  $\propto f^{-1}$.  Whether the intrinsic source of the
  variability  is the disk (Payne 1980\nocite{1980ApJ...237..951P}; Miyamoto et al. 1988\nocite{1988Natur.336..450M}) or the  corona 
(Haardt et al.~1997\nocite{1997ApJ...476..620H}, Poutanen \& Fabian 1999) is still a matter  of
  debate.
  As the characteristic time scales are expected to grow linearly with 
the black hole mass, the time scales are far longer in Seyfert galaxies.
  The long observations required prevent the acquisition of as many
 data as for X-ray binaries.
  Their Power Density Spectra (PDS), at least, are similar to those of stellar black holes, modulo the mass scale factor (Edelson \& Nandra 1999).

  To understand these temporal characteristics it
 seems important to introduce the temporal dimension in spectral models.   
  The most important  difficulty when dealing with time dependent  problems
  is that the  system is not  necessarily  in  radiative  equilibrium.  The
  corona is strongly  coupled with radiation, its physical  parameters such
  as temperature  and optical depth can fluctuate in response to changes in
  the photon  field.  These  variations  in turn  influence  the  radiative
  field, making the system strongly non-linear.
  The dynamics of compact plasma has been  extensively studied 
  during
  the eighties  in the context of  the models for the high energy
  emission of active galactic nuclei;
  first using analytical arguments  (Guilbert et al. 1983\nocite{1983MNRAS.205..593G}), then using more and more accurate numerical methods  (Guilbert
 \& Stepney 1985;\nocite{1985MNRAS.212..523G}
 Fabian et al. 1986\nocite{1986MNRAS.221..931F}; Kusunose 1987;\nocite{1987ApJ...321..186K} Done \& Fabian 1989\nocite{1989MNRAS.240...81D}). The most 
detailed treatment of micro-physics was achieved by Coppi (1992) 
\nocite{1992MNRAS.258..657C} using a method based on the solution
 of the kinetic equations. These studies gave an understanding of the 
behaviour of the plasma and the spectral evolution of the emitted 
radiation, when the input parameters (heating, external soft-photons injection...) vary on time scales of the order of the light crossing time.

  However, at the time of those studies the importance 
  of the coupling with cold matter did not appear as crucial as it does now.    
  Indeed, in the accretion disk corona framework, another complication 
  appears: the hard X-ray radiation is produced
  by  Comptonisation  of soft  photons  that, in  turn,  can be  mostly
  produced by  reprocessing  the same hard  radiation in the cold accretion
  disk  (the   ``feedback''   mechanism). 
  Until now, in most studies only steady-state  situations have
  been  considered.  The physical  characteristics  of the emitting  region
  (such as temperature  and optical depth of the  Comptonising  cloud) that
  determine the observed X/$\gamma$-ray spectrum are assumed not to vary in
  time (e.g. Sunayev \& Titarchuk 1980)\nocite{1980A&A....86..121S}.
 The most detailed calculations
considered only steady states, where the temperature and the optical depth are defined by the energy  balance
  and electron-positron  pair balance, assuming a constant heating
 (Haardt \& Maraschi 1993; Stern et al. 1995a,b\nocite{1995ApJ...449L..13S}\nocite{1995MNRAS.272..291S}; Poutanen \& Svensson 1996\nocite{1996ApJ...470..249P}).

  The observed rapid  spectral changes
  imply the presence of rapid changes in the physical  conditions  of
  the source.  When taken into  account in  radiative  transfer  modeling,
  these  changes  have been  considered as a  succession of steady state
  equilibria (Haardt et al. 1997;\nocite{1994ApJ...432L..95H} Poutanen \& Fabian 1999). This  approximation  is acceptable as long as the underlying
  perturbation  evolves on time  scales,  $t_{\mathrm{c}}$, far larger than the light
  crossing  time.  Actually, we do not know if this
  assumption is valid.

  Here we aim at giving a first look at the rapid spectral and temporal 
  evolution of a hot plasma  
  coupled with a reprocessor.
 In this first attempt to model the non-linear behavior of 
a time dependent accretion disk corona system, we try to point out  
the main properties of the accretion disk corona
 when the dissipation 
parameters change on time scales of the order of the light crossing time of 
the corona.
  We show that the non-linear Monte Carlo method (Stern et al. 1995a)
 can be an efficient tool to perform the task of computing the time evolution
of the plasma parameters together with a detailed radiative transfer treatment
enabling the production of light curves.   
\nocite{1999ApJ...514..682E}
 Our model assumptions are presented
 in Sect.~\ref{assumptions}.

Section ~\ref{montecarlo} gives a 
description of our computational method. The different consistency tests that we performed in order to check the validity of the code for steady state situations are presented in Sect.~\ref{comparisons}.
Then we present some applications to time dependent situations.
 We first investigate the case of an equilibrium modified by 
a variation in the soft photon input,  in Sect.~
\ref{variabilitydrivenbythedisk}.
 In Sect.~\ref{variabilitydrivenbythecorona}, we then consider 
situations where the energy dissipation in  the corona occurs during short   
 flares.

\section{Model assumptions \label{assumptions}}

\subsection{The slab corona model}
We consider a simple slab geometry for the corona.
 The corona is composed of electrons (associated with ions) with a fixed
 Thomson optical depth $\taus$. This corona is assumed to be uniformly heated by an unspecified process, which is  quantified using
 the usual local dissipation parameter $l_{\mathrm{c}}=H F_\mathrm{h}\sigmath/\me c^{3}$,
where  
$F_\mathrm{h}$ is the power supplied in the corona per unit area, 
$H$ the height of the slab, $\sigmath$ the Thomson cross section,
 $\me$ the electron rest mass and $c$ the speed of light. 

The corona cools by Comptonisation of the soft radiation from the disk.
 The balance between heating and cooling leads to a mean temperature $T_{\mathrm{e}}$.
 Due to photon-photon interactions, $e^{+}/e^{-}$ pairs can appear 
in the corona leading to a total optical depth $\taut>\taus$ that is 
governed by the balance between pair production and annihilation.  

The disk emission arises from two processes:

\begin{itemize}
\item  internal viscous dissipation parametrized by the disk compactness 
$l_{\mathrm{d}}=HF_\mathrm{s}\sigmath/\me c^{3}$,
 where  $F_\mathrm{s}$ is the intrinsic flux of radiation emitted by the disk 
with a blackbody spectrum at fixed 
temperature $T_{\mathrm{bb}}$. 

\item  reprocessing of Comptonised radiation coming from the corona.
 Most of the impinging radiation is absorbed and thermally re-emitted 
with blackbody temperature $T_{\mathrm{bb}}$, a fraction is Compton reflected forming 
a hump in the high energy spectrum. 
\end{itemize}

\subsection{Steady state properties}

The equilibrium properties of accretion disk corona have been extensively
studied  (Haardt \& Maraschi 1993; Stern et al. 1995a, b;\nocite{1995ApJ...449L..13S}\nocite{1995MNRAS.272..291S} Poutanen \& Svensson 1996;
\nocite{1996ApJ...470..249P} Dove et al. 1997b\nocite{1997ApJ...487..747D}\nocite{1997ApJ...487..759D}). Let us recall their main characteristics.

For a given total optical depth $\taut$, the plasma temperature and thus the spectral properties depend only on
the ratio $l_{\mathrm{d}}/l_{\mathrm{c}}$ and are independent of the absolute luminosity.
When $l_{\mathrm{d}}/l_{\mathrm{c}}\gg1$, heating is negligible, the plasma
is at Compton temperature driven by the disk thermal  radiation $T_{\mathrm{e}}\sim T_{\mathrm{r}}$.
Decreasing  $l_{\mathrm{d}}/l_{\mathrm{c}}$ quickly raises the temperature provided that 
$l_{\mathrm{d}}/l_{\mathrm{c}}>0.1$. For smaller values of the ratio $l_{\mathrm{d}}/l_{\mathrm{c}}$, the temperature becomes independent of 
$l_{\mathrm{d}}/l_{\mathrm{c}}$. It is easily understandable since the cooling is then dominated by 
the reprocessed photons whose energy density grows linearly with the heating 
rate $l_{\mathrm{c}}$. For a given optical depth
there is thus a maximum self-consistent temperature. This maximum 
temperature depends only on the feedback coefficient which is defined as the
fraction of the X-ray flux which reenters the source as soft radiation 
after reprocessing. This feedback coefficient depends mainly on the geometry.
Thus for a given geometry the hardness of the emitted spectra is limited by the maximum temperature achievable.
  
Based on these arguments, it has been often argued that the slab geometry produces 
too steep spectral slopes to account for the observed spectra in black hole binaries and Seyfert galaxies (Stern et al. 1995b; Dove et al. 1997b; Poutanen et al. 1997).\nocite{1995ApJ...449L..13S}\nocite{1997ApJ...487..759D}\nocite{1997MNRAS.292L..21P} The geometry is more likely a spherical corona surrounded by the cold accretion disk, or constituted of small active regions
at the surface of the disk.  In these cases indeed the feedback coefficient is lower. Note, however, that slab corona could be consistent with observations if the disk is ionised
 (Nayakshin \& Dove 1999; Ross et al. 1999)\nocite{1999MNRAS.306..461R}\nocite{nayak99} or if there is a bulk motion of plasma away from the disk (Beloborodov 1999).\nocite{1999ApJ...510L.123B}

Decreasing $\taus$ raises the temperature since the injected energy 
is shared by a smaller number of particles.
If $l_{\mathrm{c}}/\taus>1$ and $l_{\mathrm{d}}/l_{\mathrm{c}}\ll 1$, the pair production can increase 
significantly the total optical depth $\taut$. Then the equilibrium optical
 depth has to be computed using the pair production and annihilation balance 
and will depend on the compactness $l_{\mathrm{c}}$. An increase in $l_{\mathrm{c}}$ 
increases the pair production rate and so $\taut$ which in turn decreases 
the temperature.

\subsection{Time dependent situations}

Here we will 
consider rapid changes in  $l_{\mathrm{c}}$ or $l_{\mathrm{d}}$. The dissipation parameters
  $l_{\mathrm{c}}$ and $l_{\mathrm{d}}$ are thus functions of time with a characteristic time scale of a few $H/c$.

Even if the simple slab geometry considered here seems rather unrealistic 
with regard to the observations, we do not expect that the general features 
presented here change qualitatively for a different geometry.

To simplify our calculations,  we made several approximations discussed below:

First, we assume thermal electron distributions. It is well known that if the 
temperature changes are too fast the particles may not have time
to form a true Maxwellian distribution. The main consequence is the 
formation of a high-energy tail in the electron distribution (Li et al. 1996\nocite{1996A&AS..120C.167L}; Poutanen \& Coppi 1998\nocite{1998Poutetcop})
that could
 explain the emission observed at gamma-ray energies in black hole candidate Cygnus X-1 (Ling et al. 1997\nocite{1997ApJ...484..375L}; McConnell et al. 1997\nocite{1997comp.symp..829M}). Unless the medium is very optically thin, the hard-X ray spectrum should not be significantly affected. Indeed, the Comptonisation spectrum is 
 mainly sensible to the mean electron energy and not to the detailed 
particle distribution when multiple scattering is important 
(Ghisellini et al. 1993\nocite{1993MNRAS.263L...9G}).

We also neglect the time delays due to the radiative transfer in the disk.
The reprocessing and reflection are supposed to be instantaneous.
The travel time for a photon in the disk scales as the inverse of the disk density.
 The disk being dense  and optically thick,
 the time delays due to 
light traveling time in the disk are negligible compared to the corona
 light crossing time. Concerning the reprocessed component the delays are
 mainly due to the thermalisation time of the disk which is short  (typically $\sim10 \mu$s, see e.g. Nowak~et~al.~1999)\nocite{1999ApJ...515..726N} compared to the corona light crossing time.

Another important simplification is that the blackbody temperature of 
the disk is fixed and considered to be constant. We can expect that the 
disk temperature will have important fluctuations (scaling as $l^{1/4}$) 
as a response to changes
 in the illuminating flux or the intrinsic dissipation parameter.
Note however that  we take into account the disk 
luminosity variations in a self-consistent manner. Our specification of a fixed temperature fixes
 only the disk spectrum shape and not its amplitude.  

We thus expect that calculations including the temperature changes 
would give different results only for the spectral evolution in the 
soft X-ray
 bands ($<$ 1 keV). At higher energy where the flux is dominated by Compton emission,
 the results should not be affected. Indeed, at first order, the Compton losses scale as the soft photon energy density rather than photon energy.  
 We checked that the temporal evolution is not qualitatively sensible
to the value of the fixed blackbody temperature, which can fluctuate within a factor of 10 without changing significantly the high-energy light curves.
For detailed quantitative considerations, however, these effects
will have to be implemented in the code.

\section{Monte-Carlo Code\label{montecarlo}}
To compute the evolution of $T_{\mathrm{e}}$ and $\taut$ together with the flare light
 curves, we use a Non-Linear Monte-Carlo code (NLMC) that we developed 
according to the Large Particle (LP) method proposed by Stern et al. (1995a)
\nocite{1995MNRAS.272..291S}.
 The main features of our code are similar to Stern's.
 The radiative processes taken in account are Compton scattering, 
pair production and annihilation. 
A pool is used to represent the thermal electron population. 
 The reflection component is computed with a coupled linear code 
(Jourdain \& Roques 1995;\nocite{1995ApJ...440..128J} Malzac et al. 1998)\nocite{1998A&A...336..807M}.
 We implemented in our NLMC code the slab corona described in Stern et al (1995b) (see also Dove et al. 1997a).
 The slab is divided into ten homogeneous layers to account for its vertical
structure.

 An ample discussion of the LP method can be found in Stern et al. (1995a,b) and further details on its application to thermal accretion
 disk corona models can be found in Dove et al. (1997a), we only discuss here the aspects of our code which are related to temporal variability.

Unlike standard Monte-Carlo Methods (Pozdnyakov et al. 1983\nocite{1983SSRvE...2..189P}; Gorecki \& Wilczewski 1984\nocite{1984AcA....34..141G}), the LP method propagates all the particles
 in a parallel and synchronized way.
 Thus, the time variable appears as a natural parameter. 
However, when dealing with time-dependent (TD) systems the problem 
of statistical errors becomes crucial. Indeed, when simulating a stationary system
 the spectrum has just to be integrated over a longer time (in LP method)
 or over more particles (in a standard MC method) to get the required
 statistical accuracy.
 In TD situations this is of course not the case and the statistical accuracy
 depends on the number of LP, $N_{\mathrm{LP}}$, used to represent the system 
at a given time. Due to the weighting 
technique used, we do not require a very large number of LPs to
 get a statistically acceptable representation of the particles distribution
 \emph{inside} the active region (here we use 5~10$^4$ to 15~10$^4$ LPs).
 However, the number of \emph{escaping} photon LPs per time step 
 is only a very small fraction of $N_{\mathrm{LP}}$ 
($\sim\Delta tN_{\mathrm{LP}}$, $\Delta t=10^{-3} H/c$ here).
 This prevents the averaging of light curves over too short time scales or too
 narrow energy ranges. 

 It is difficult to improve accuracy by increasing the number of LPs, $N_{\mathrm{LP}}$. Indeed the statistical errors scale roughly as
$1/\sqrt{N_{\mathrm{LP}}}$. Thus, in order to decrease these errors by a factor of 2, 
$N_{\mathrm{LP}}$ has to be increased by a factor of 4. 
The simulation time can be estimated scaling as $N_{\mathrm{LP}}\log{N_{\mathrm{LP}}}$ 
(Stern~et~al.~1995a), the performances are thus degraded
 by more than a factor of 4.
This is not negligible particularly when the CPU time is already large (as large as 1 day).

 However in our case,
 we are mainly interested in the average
 spectrum during a flare, and the light curves averaged over $0.1H/c$ and 
a decade in energy are sufficient to get 
the general properties of the flare.

Another problem due to the statistical method is the determination
 of the electron pool parameters $T_{\mathrm{e}}$ and $\taut$. We have
to calculate the pool energy  $\Delta E_{\mathrm{p}}$ and optical depth changes
 $\Delta\taut$ during $\Delta t$.
 If this is done using the Monte-Carlo interactions that occurred during the 
time step, we get huge statistical fluctuations. 
To limit these fluctuations, Stern et al. (1995a) average the temperature over previous
 time steps.
 This method, rigorous only  for equilibrium states, introduces in TD simulations an artificial relaxation time.
 We overcome the problem by computing $\Delta E_{\mathrm{p}}$ and $\Delta\taut$
 analytically.
We  use the exact thermal annihilation rate given by Svensson (1982)\nocite{1982ApJ...258..321S} and 
the formula given by Barbosa (1982)\nocite{1982ApJ...254..301B} and
 Coppi \& Blandford (1990) \nocite{1990MNRAS.245..453C} to compute and
 tabulate the exact energy exchange rate between a photon of given energy 
and a Maxwellian distribution of electrons through the Compton process.
 These results are used at the beginning of each time step to evaluate 
the Compton losses using the updated distribution of photon LPs. 
This method allows us to limit the step to step fluctuations to less than 1 
per thousand.
 However, statistical fluctuations of the photon distribution can lead
 to fluctuations of 5 $\%$ on a time scale of a few $H/c$.
 Fortunately, these oscillations occur only in stationary situations where
 the electron temperature is more sensitive to photon LP fluctuations.
 In non-equilibrium situations the temperature is strongly driven toward
 equilibrium and evolves smoothly.  

\begin{figure}[t]
\centerline{
\scalebox{1.}{\includegraphics{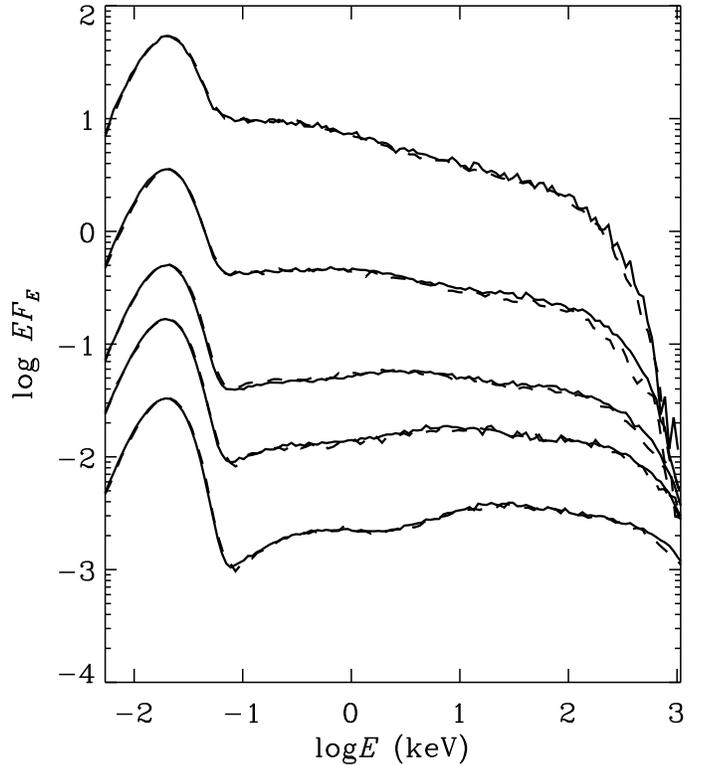}}}
\caption{Comptonisation spectra emitted by  a slab corona
 in energy balance. Disk blackbody temperature $kT_{\mathrm{bb}}$ is 5 eV. Spectra are
 averaged over inclinations $\theta$ such that $0.6<\cos\theta<0.9$, where 
$\theta$ is the angle between the line of sight and the disk normal.
 Solid lines 
show spectra computed with  our code while  Stern's results are shown in 
dashed lines. Fixed optical depths are, from bottom to top  $\taut=$0.033, 
0.058, 0.090, 0.163 and  0.292.
The equilibrium temperatures we found are respectively $kT_{\mathrm{e}}=$425, 
300, 225, 152
 and 98 keV. Stern's values for equilibrium temperature are very close
 to the latter, with the most important difference at $\taut=0.163$, for
 which his temperature is 3$\%$ lower. 
\label{comparstern}}
\end{figure}

\section{Validation of the code\label{comparisons}}

\subsection{Emitted spectra and energy balance}

To check that our code produces accurate Comptonisation spectra,
we switched off pair production and annihilation processes 
and compared our results to spectra from a linear code 
(see, e.g., Pozdnyakov et al., 1983).
In these tests, the energy balance was not considered, optical depth and 
temperature were fixed.
The emergent-angle-dependent spectra as well as the energy and angular distributions of radiation inside the active region, are similar within 
statistical errors (a few percent). These tests have been performed using
 a pool to represent the electron distribution. We performed  other successful tests when  electron LPs are drawn at the beginning of the time step
 over a fixed distribution (Maxwellian or power law) and
for several geometries (sphere, cylinder, slab).    
 Similar comparisons have been performed in the case of a corona coupled with a
reprocessor (including the reflection component). Here again a very good agreement was found.
 
For a slab corona coupled with a disk, taking into account the energy 
balance provides, for a given optical depth, an equilibrium temperature which 
can be compared with results from the literature. Stern et al. (1995b) and Poutanen \& Svensson (1996) give maximum temperatures which are in agreement with
 ours within 5 \%. Emitted spectra are also in agreement as shown 
in Fig~\ref{comparstern} which compares spectra from our code against results from Stern's, for the same fixed optical depth, and a temperature determined
 according to the energy balance.

In addition we tested the energy conservation,
by checking that the whole luminosity emitted by the corona+disk system 
equals the injected power.

These tests validate the treatment of Compton scattering as well as the
 general architecture of the code and all the routines that do 
not depend on the type of interaction (LPs management, geometry...),
 i.e. the main parts of the code.

\subsection{Pair production and annihilation}

The pair annihilation rate as well as the spectra of produced photons have been compared, in the thermal case, to the analytical formulae given by Svensson (1982) and Svensson et al. (1996)\nocite{1996A&AS..120C.587S}. The differences obtained are less than 1 \%.

The pair production rate is more difficult to test because it is strongly sensitive to the number density, energy and angular distribution
 of the photons around and above the electron rest mass energy, $m_{\mathrm{e}}c^{2}$.
 The pair production rate thus depends strongly on the details of 
radiative transfer such as the number and shape of spatial cells used (the radiation field being averaged over a cell).

The pair production rate obtained with our NLMC code has been compared with a pair production rate computed analytically from the  photon LP distribution
 provided by the same simulation. Both pair production rates are in agreement within statistical fluctuations (a few percents).

However, detailed comparisons with the results of Stern et al. (1995b) in the framework of accretion disk corona models show important differences for the pure pair plasma case when the optical depth (and compactness) is low ($\taut<0.1$). This disagreement leads to differences in optical depth (and thus temperature) up to 25\% for $\taut\sim10^{-2}$ ($l_{\mathrm{c}}<1$). The origin of these differences 
is still not clear; it could be due to differences in the details of the 
photon-photon interaction treatment, or to the use of a different number 
of zones, or simply to statistical errors. Indeed Stern et al.~(1995a) 
used 5 times less LPs than we did.
However, for less extreme parameters ($l_{\mathrm{c}}>1$), the equilibrium parameters differ by more than 5 \% only occasionally.  
  We also tested our equilibrium values against the results presented in the inset of Fig.~2 of Dove et al. (1997a), ($\taus=0.2$, $l_{\mathrm{d}}=1$ and  $1<l_{\mathrm{c}}<100$),  a perfect agreement was found.
In addition we performed comparisons with the ISM code (Poutanen \& Svensson 1996) for the case of pure pair plasmas in pair and energy balance 
in slab geometry and found resulting  equilibrium optical depths and temperatures differing by less than 10\%.

\section{Variability driven by the disk\label{variabilitydrivenbythedisk}}

\subsection{Set up}

  The first class of models for the rapid aperiodic variability of 
 GBHC uses the variability of the 
soft seed photons as a source for the hard X-ray variability  
  (e.g., Kazanas et al. 1997; Hua et al. 1997; B\"ottcher \& Liang~1998).
\nocite{1997ApJ...480..735K}\nocite{1997ApJ...482L..57H}
\nocite{1998ApJ...506..281B}
 Soft  photons  are
  assumed to be  injected  isotropically  in the centre of a very  extended
  corona  with a white  noise  power  spectrum. The Compton process wipes out 
  the high frequency variability giving the overall shape of the PSD, and 
the hard lags are due to the soft photons Compton up-scattering time.  
Another attractive model of this kind has been recently proposed by
 Böttcher \& Liang (1999)\nocite{1999ApJ...511L..37B} based upon an idea previously quoted by 
Miyamoto \& Kitamoto (1989)\nocite{1989Natur.342..773M}.
In this scheme it is  assumed that the soft seed photons are emitted by cool dense blobs 
of matter spiraling inward through an inhomogenous spherical corona formed by
 the inner hot disk.
These models have however an intrinsic deficiency: they do not
 account for the energy balance.  The comptonising plasma temperature is assumed to
 be constant while  
 the soft photon field 
changes very quickly in time. As the Compton scattering on soft photons
 is the dominant cooling mechanism for the thermal electrons, the temperature is
 determined mainly by the soft photon energy density $U_{\mathrm{s}}$ and the
 heating rate. Any 
modification of the soft photons field induces change in temperature.
 The cooling time for the thermal electrons is very short, scaling roughly
 as $1/U_{\mathrm{s}}$. Thus the temperature 
adjusts very quickly to any significant change in the soft radiation field.
 These temperature fluctuations in turn affect the emitted spectrum.
 It is then important to know what really happens when 
the temperature equilibrium is  perturbed by a varying soft photon input.

In our slab corona we consider soft photon flares arising from the disk. We model a flare  by a sudden increase in the
disk internal dissipation parameter $l_{\mathrm{d}}$. 

\subsection{Results of simulations}

\begin{figure}[tb]
\centerline{
\scalebox{.5}{\includegraphics{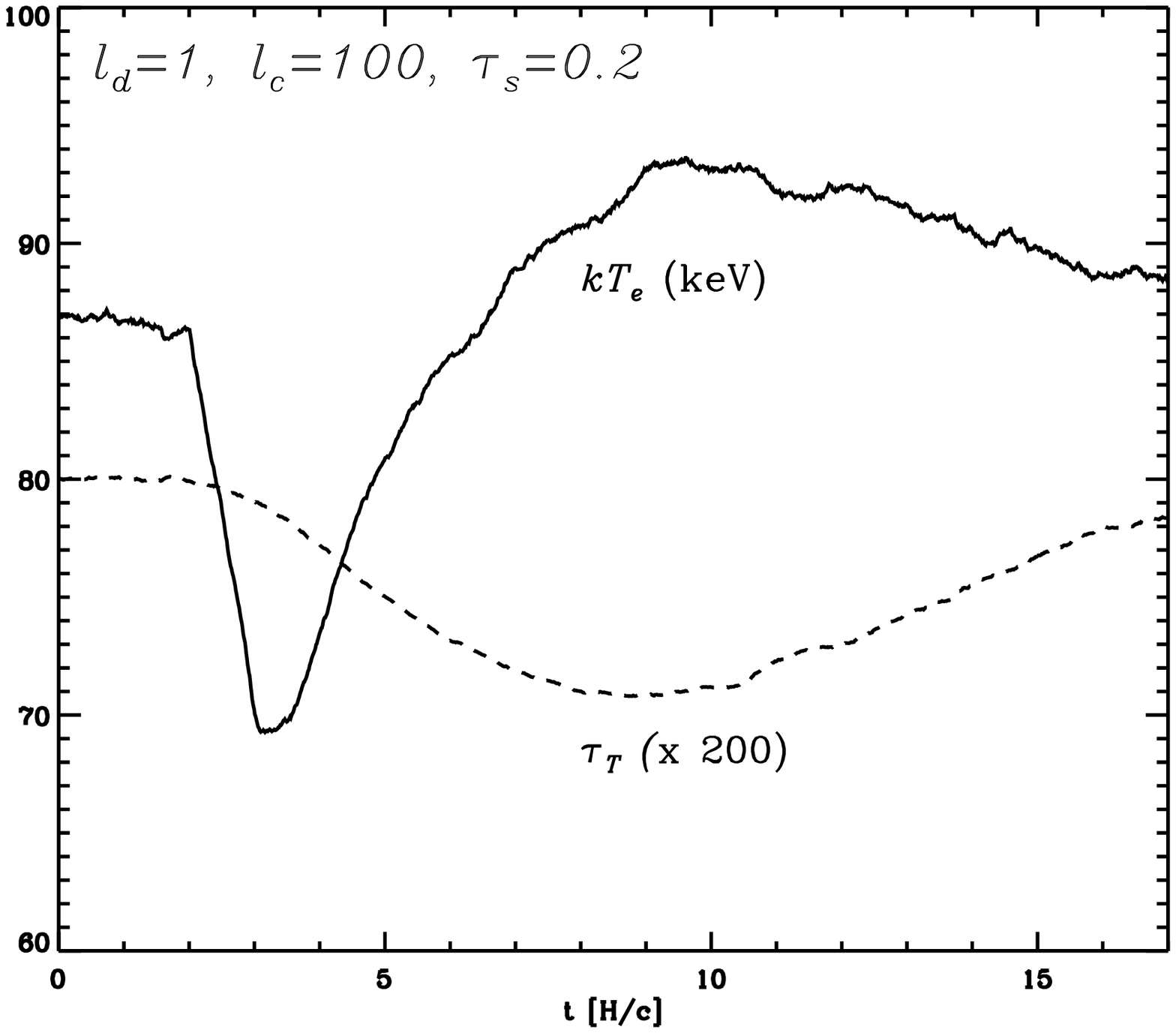}}}

\centerline{
\scalebox{.5}{\includegraphics{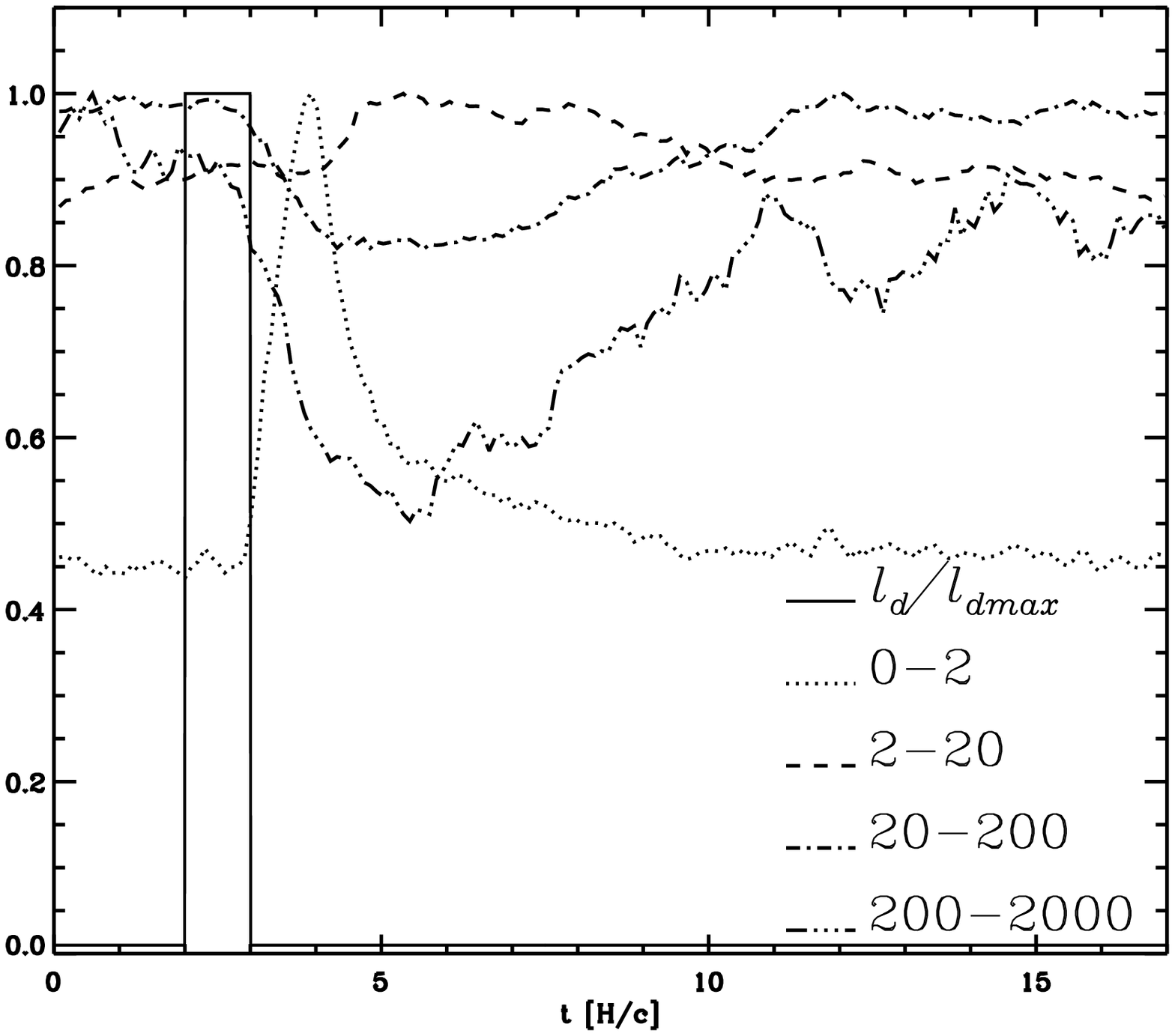}}}
\caption{Upper panel: Evolution of the mean coronal temperature (solid line) and optical depth (dashed line) as a response to a flare in the disk.
 The system is initially at equilibrium with $\taus=0.2$, $l_{\mathrm{c}}=100$, $l_{\mathrm{d}}=1$, $kT_{\mathrm{bb}}$=200 eV. Between $t=2$ and $t=3$, $l_{\mathrm{d}}=100$,
Lower panel: 
The profile of the disk dissipation parameter $l_{\mathrm{d}}$ is shown in solid line,
together with the light
 curves in the 0-2 keV, 2-20 keV, 20-200 keV, 200-2000 keV. 
All
 curves 
are 
normalised 
to their maximum.\label{disk1}
}
\end{figure}

Fig.~\ref{disk1} presents the evolution of the coronal physical parameters
in the case of an equilibrium perturbed by a strong and violent emission of
soft photons from the disk.

 The system is initially in a steady 
state where the internal disk dissipation parameter  $l_{\mathrm{d}}=1$ 
while $l_{\mathrm{c}}=100$.
 $l_{\mathrm{d}}$ is then increased by 
a factor of 100 during $\Delta t=1$ $H/c$ (hereafter time is expressed in $H/c$ units). 
The Compton cooling 
of the plasma is quasi-instantaneous and the temperature drops by $\sim15\%$.
 The pair production rate thus 
decreases, leading to a lower optical depth.
After perturbation, the optical depth increases again, but slower than the 
lepton kinetic energy.
Thus the temperature increases slowly and reaches 
a maximum higher than the initial equilibrium temperature. This arises from the pair production time being longer than the heating time.  
After the event the system  relaxes toward equilibrium.

The code also enables us to compute the associated light curves (Fig.~\ref{disk1}).
The soft luminosity in the lower energy band ($E<$2 keV) increases strongly 
by a factor of $\sim 2$ around 1 $H/c$ after the beginning 
of the flare. Note however that this is a small variation compared to the
 change of a factor 100 in $l_{\mathrm{d}}$ that we imposed. Indeed in the steady state
 the disk emission is the sum of intrinsic dissipation and reprocessing of
 hard radiation.

The delay is due to the corona light crossing time. The flux in the highest 
energy bands (20-200 keV and 200-2000 keV) decreases due to 
the temperature drop.
 On the other hand, the flux in the intermediate band (2-20 keV) increases slightly.

 However
 the overall Comptonised radiation flux is roughly constant, since the dissipation 
in the corona is kept constant. \emph{If $l_{\mathrm{c}}$ is constant, a modification of the seed photon flux does not change the integrated luminosity in the 1-2000 keV band}. The temperature adjusts very quickly to maintain constant Compton losses.
The spectrum thus appears to pivot around $\sim$ 20 keV.

Note however that the variations in the energy range usually used in observations (2-50 keV) are very weak (at most  20 \%), far lower than those usually observed in X-ray binaries (RMS$\sim 30 \%$). 

\begin{figure}[tbh]
\centerline{
\scalebox{.5}{\includegraphics{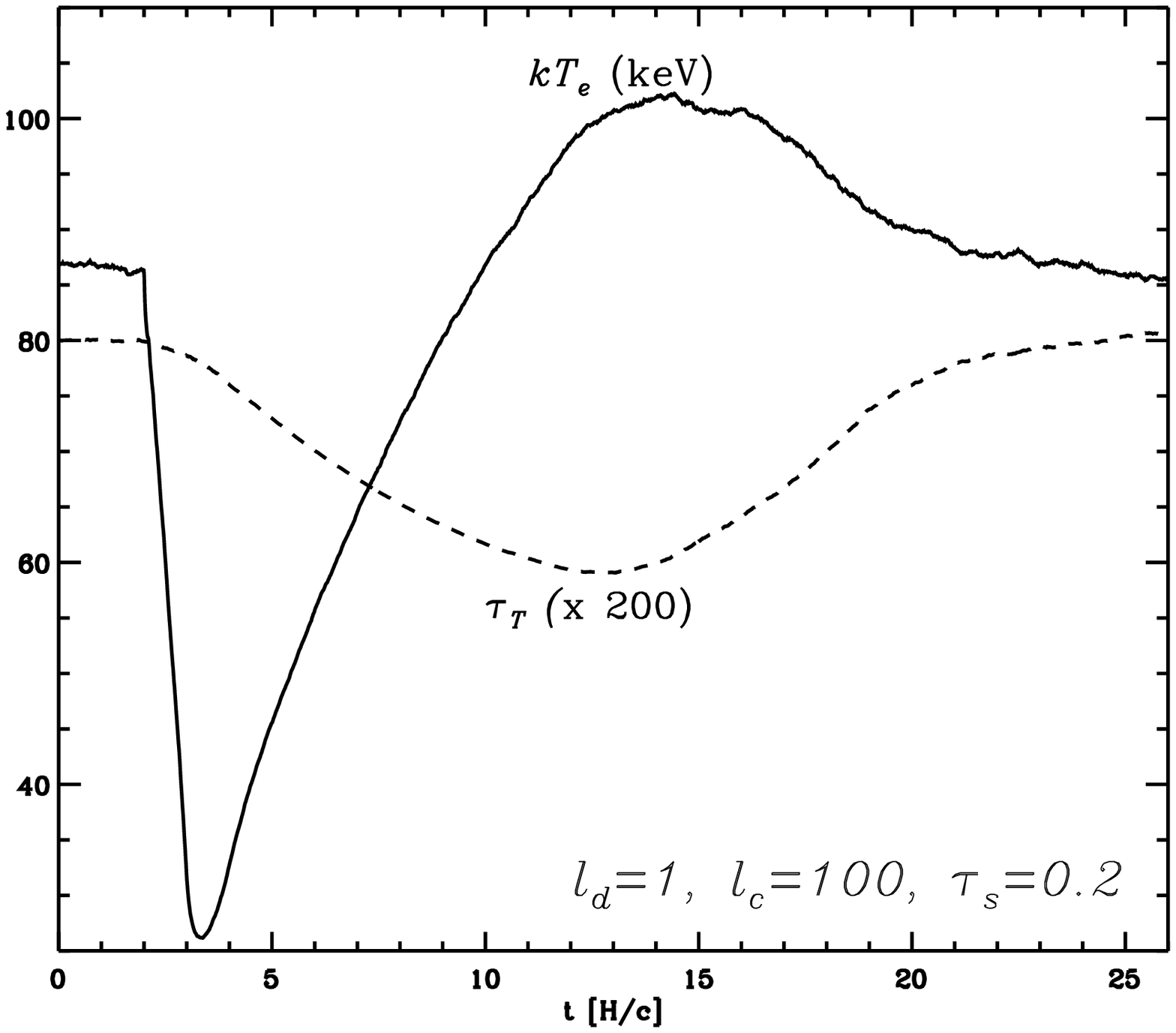}}}

\centerline{
\scalebox{.5}{\includegraphics{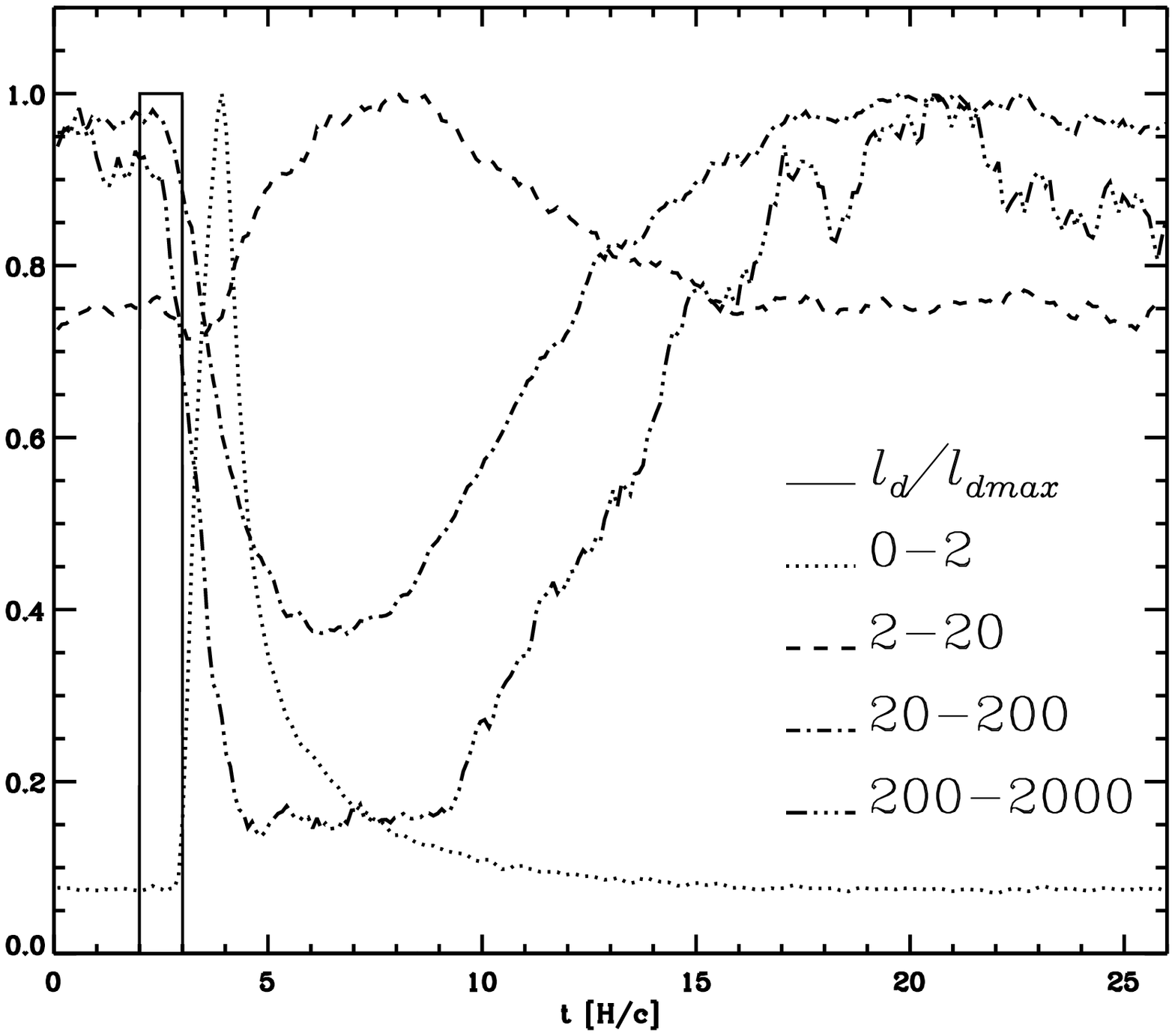}}}
\caption{Upper panel: Evolution of the mean coronal temperature (solid line) and optical depth (dashed line) as a response to a flare in the disk.
 The system is initially at equilibrium with $\taus=0.2$, $l_{\mathrm{c}}=100$, $l_{\mathrm{d}}=1$, $kT_{\mathrm{bb}}$=200 eV. Between  t=2 and t=3, $l_{\mathrm{d}}=1000$.
Lower panel: 
The profile of the disk dissipation parameter $l_{\mathrm{d}}$ is shown in solid line,
together with the light
 curves in the 0-2 keV, 2-20 keV, 20-200 keV, 200-2000 keV. 
All curves are normalised to their
 maximum.\label{disk2}
}

\end{figure}

Higher amplitude fluctuations can be obtained for a larger perturbation
 amplitude. For example, Fig.~\ref{disk2} displays the temperature and optical depth evolutions for $l_{\mathrm{d}}=1000$ between $t=2$ and $t=3$, instead of $l_{\mathrm{d}}=100$
in the previous example, the other parameters being unchanged. 
As the cooling effect is now stronger the temperature drops by $\sim 80$\%.
 The spectral evolution is important, the light curves, shown in Fig.~\ref{disk2},
display significant fluctuations in the four energy ranges. 

Important fluctuations can also be obtained by increasing the shot
 characteristic timescale. Longer durations indeed enable to reach the low
 temperatures obtained at equilibrium for large $l_{\mathrm{d}}$. 
 
By varying the shot timescale and amplitude we can thus get complicated 
correlation and anticorrelation between energy bands.
 Another complication that we do not consider here is that there 
may be a rapid succession of shots in the disk.
If  these shots are close enough in time and space, the corona has no 
time to relax toward equilibrium between each event. 
The resulting light curves  will thus also depend on the details of the 
temporal shot distribution.

\subsection{Discussion}
Current models which invoke the intrinsic seed photon variability as
 the source of the hard Comptonised radiation variability do not take  
into account the response of the corona   (Kazanas et al. 1997; Hua et al. 1997; Böttcher \& Liang 1998)\nocite{1997ApJ...480..735K}\nocite{1997ApJ...482L..57H}\nocite{1998ApJ...506..281B}. The light curves and PDS are
 computed using linear Monte-Carlo codes. The coronal characteristics 
($T_{\mathrm{e}}$ and $\taut$) are fixed.
 The high energy flux thus varies linearly with the soft photon 
flux changes. Actually, the calculations are made as if the  
heating rate in the Comptonising medium was changed according to the soft
 photon flux to compensate exactly for the Compton losses.
 We do not know how the disk is physically coupled with the corona,
 there may be some correlations between $l_{\mathrm{c}}$ and $l_{\mathrm{d}}$. It seems however
very unlikely that they adjust so perfectly to keep the temperature  constant.

Thus, any model where the variability is driven by changes in the soft photon flux
should also specify a coronal heating and take these effects into account.

Note also that the amplitude of fluctuations induced by soft photon shots 
increases with energy. Indeed the 
higher energy part of the spectrum is the most sensible to temperature 
fluctuations. This seems to be inconsistent with 
observations of X-ray binaries showing that the RMS variability is nearly independent of energy (e.g. Nowak et al. 1999)\nocite{1999ApJ...510..874N}.

On the other hand, if the shots have an amplitude and timescale long enough
 to make the corona very cool, the spectral pivot point may 
 shift toward the soft X-ray energies, leading to an anticorrelation between 
the disk and Comptonised radiation. Such a mechanism might explain 
the strange anticorrelation between UV and X-rays observed in the 
Seyfert 1 galaxy  \object{NGC7469} by Nandra et al. (1998).\nocite{1998ApJ...505..594N}

\clearpage

\section{Variability driven by the corona\label{variabilitydrivenbythecorona}}

\subsection{Set up}

In a second class of models, the observed variability is
 the consequence of rapid physical changes in the corona.
 At present the most advanced  model is that 
   of Poutanen \& Fabian (1999). 
   As in the solar corona case, magnetic
   reconnection  can lead to  violent  energy  dissipation  in the  corona.
   Accelerated/ heated particles emit  X/$\gamma$-rays by Comptonising soft
   photons  coming from the cold disk.
  The estimated  time-scales  for the
   reconnection appear to be too short to explain the fact that most of the
   power is emitted at low Fourier  frequencies.  If,  however,  the flares
   are  statistically  linked  (i.e.  each flare has a given  probability  to
   trigger one or more other flares), it can lead to long  avalanches. The
   duration of the avalanches then determines the Fourier frequencies where
   most of the power emerges. In order to explain the variability data 
this model
   requires flares with millisecond duration.  If we assume that the size of
 the corona is of the order of


\begin{figure}[tbh]
\centerline{\scalebox{1.}{\includegraphics{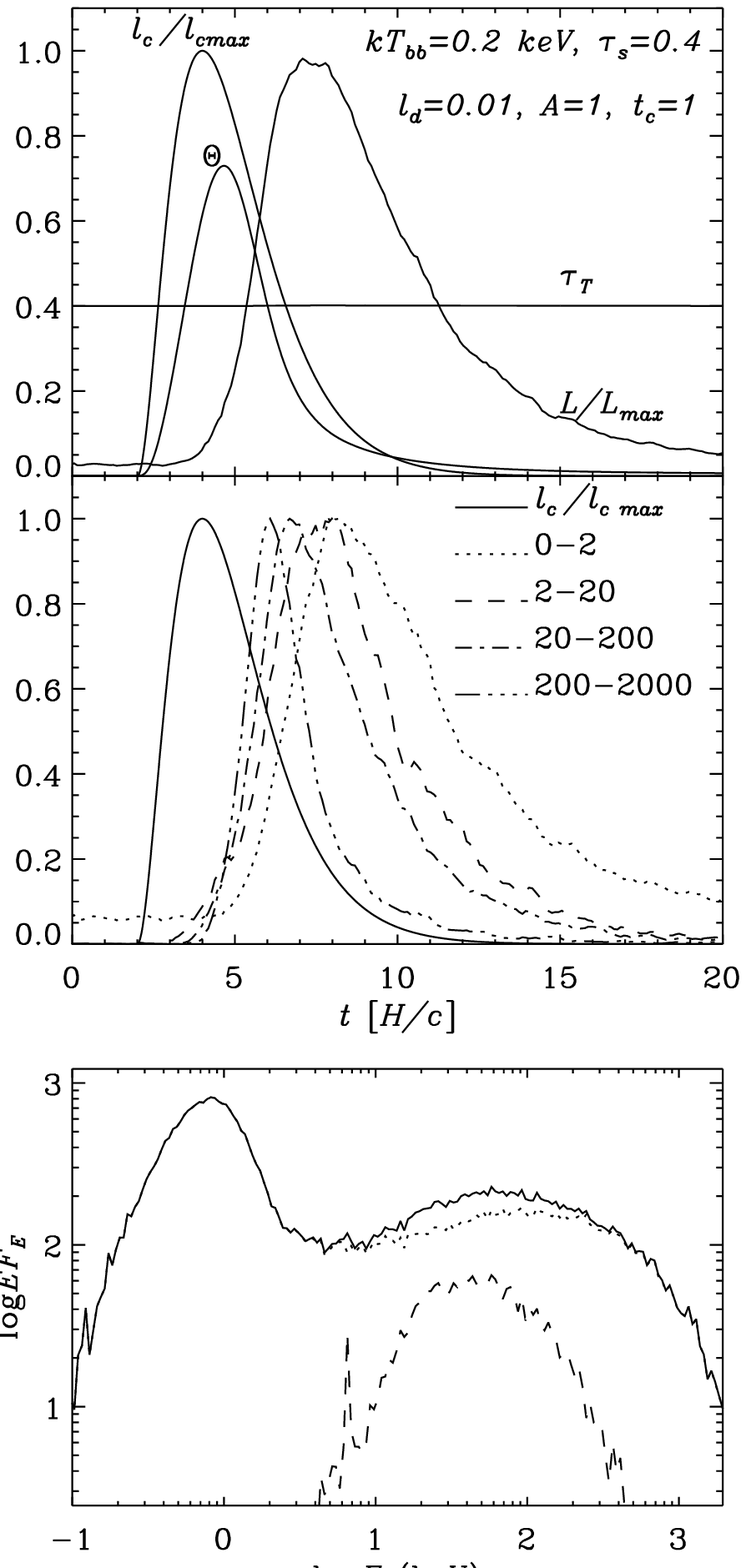}}}
\caption{Upper panel: Response of the coronal parameters
 to a strong dissipation in the corona.  
The normalised $l_{\mathrm{c}}$ profile defined by Eq.~(\ref{dissip1}) is shown together with the volume
 averaged temperature, $\theta=\kte/\mcc$, response , total optical depth, and the system total 
luminosity light curve. Middle panel: associated light curves. The profile of the disk dissipation parameter $l_{\mathrm{d}}$ is shown in solid line. Energy bands used are given in keV. All curves are normalised to their maximum. Bottom: Time average spectrum. The primary emission is shown in dotted line, the  
 reflection component in dashed line.\label{simu02}}
\end{figure}
\begin{figure}[tbh]
\centerline{\scalebox{1.}{\includegraphics{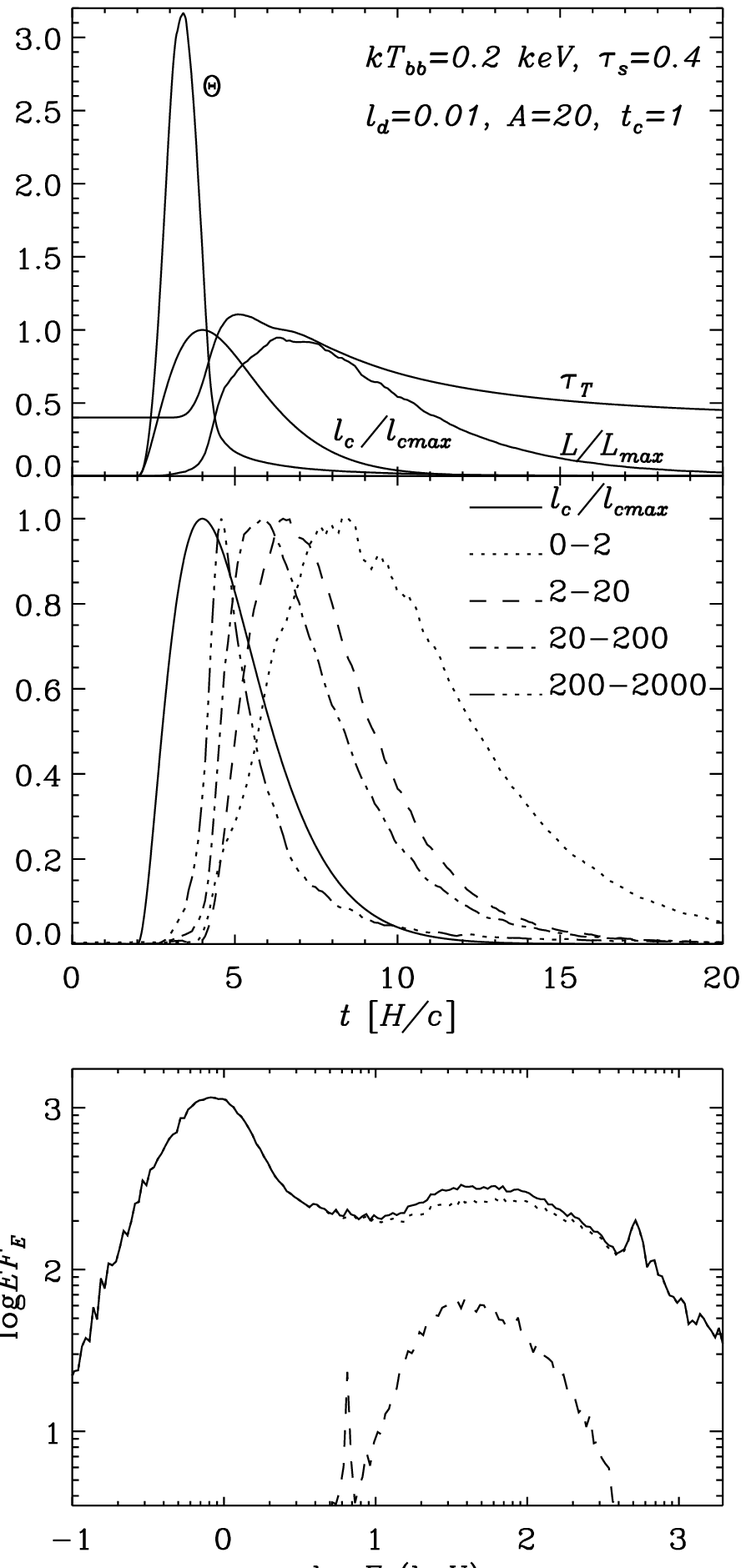}}}

\caption{Upper panel: Response of the coronal parameters
 to a strong dissipation in the corona.  
The normalised $l_{\mathrm{c}}$ profile defined by Eq.~(\ref{dissip1}) is shown together with the volume
 averaged temperature, $\theta=\kte/\mcc$, response , total optical depth, and the system total 
luminosity light curve. Middle panel: associated light curves. The profile of the disk dissipation parameter $l_{\mathrm{d}}$ is shown in solid line. Energy bands used are given in keV. All curves are normalised to their maximum. Bottom: Time average spectrum. The primary emission is shown in dotted line, the  
 reflection component in dashed line.\label{simu03}}
\end{figure}

\begin{figure}[tbh]
\centerline{\scalebox{1.}{\includegraphics{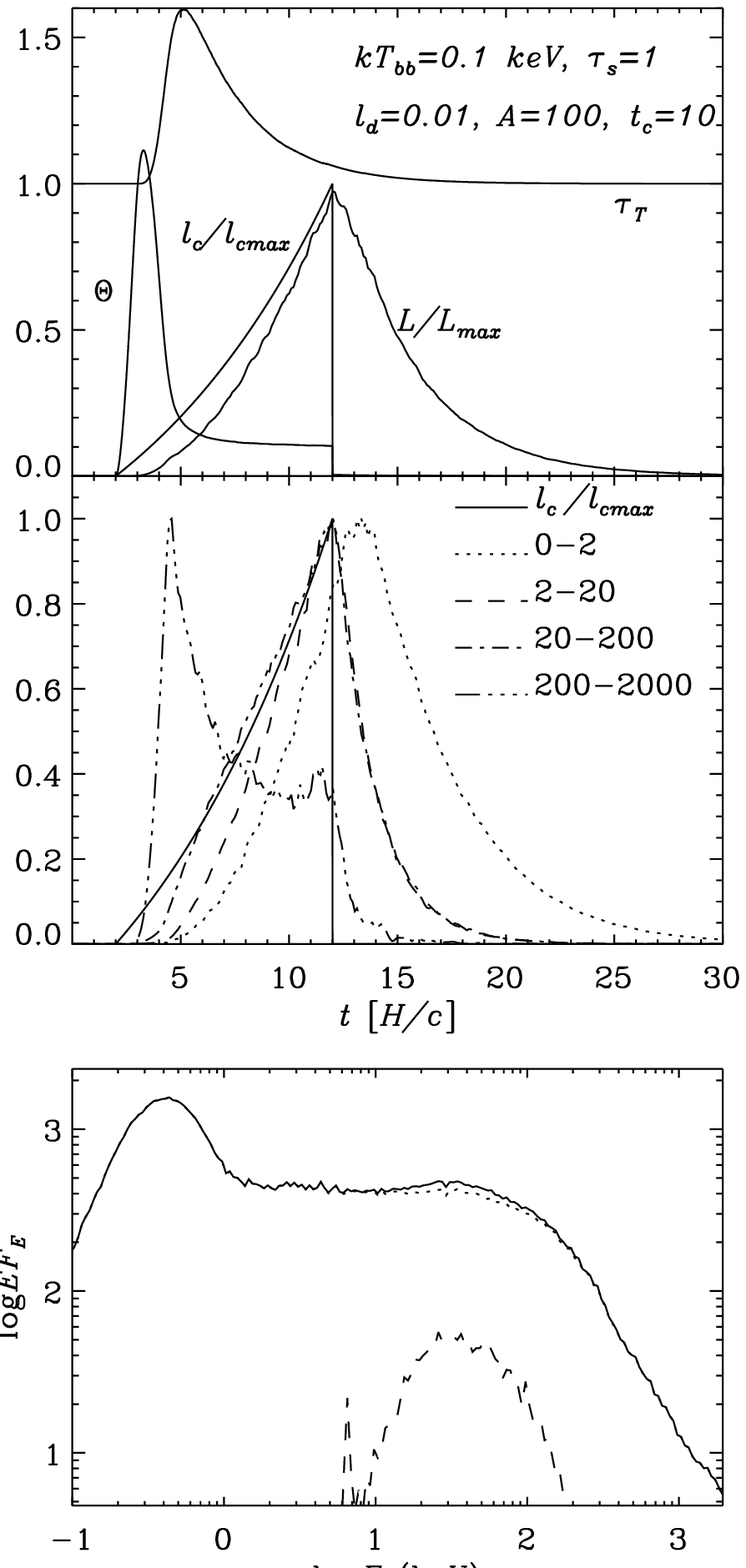}}}
\caption{Upper panel: Response of the coronal parameters
 to a strong dissipation in the corona.  
The normalised $l_{\mathrm{c}}$ profile defined by Eq.~(\ref{dissip2}) is shown together with the volume
 averaged temperature, $\theta=\kte/\mcc$, and total optical depth response, and the system total 
luminosity light curve. Middle panel: associated light curves. The profile of the disk dissipation parameter $l_{\mathrm{d}}$ is shown in solid line. Energy bands used are given in keV. All curves are normalised to their maximum. Bottom: Time average spectrum. The primary emission is shown in dotted line, the  
 reflection component in dashed line.\label{simu09}}
\end{figure}

\begin{figure}[tbh]
\centerline{\scalebox{1.}{\includegraphics{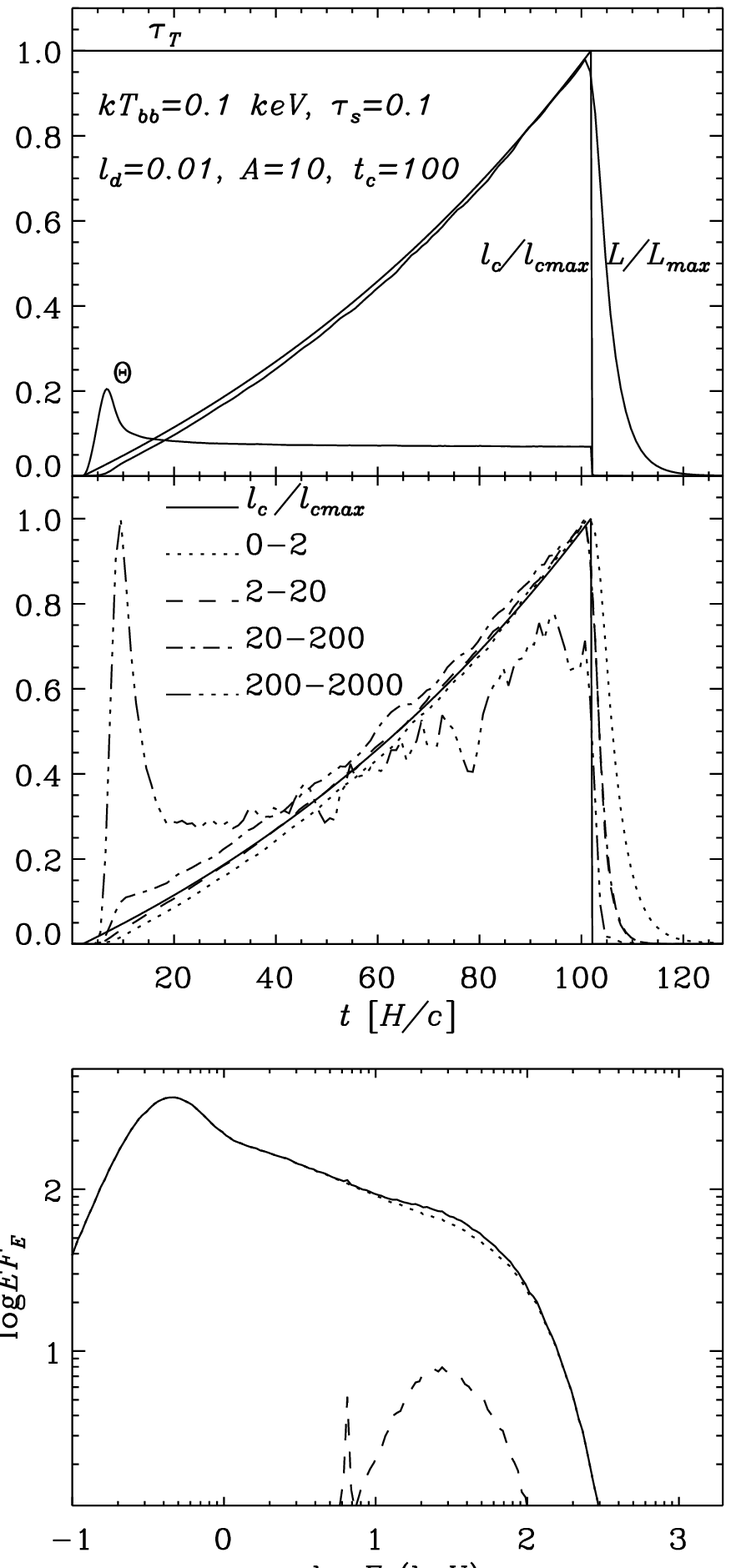}}}
\caption{Upper panel: Response of the coronal parameters
 to a strong dissipation in the corona.  
The normalised $l_{\mathrm{c}}$ profile defined by Eq.~(\ref{dissip2}) is shown together with the volume
 averaged temperature, $\theta=\kte/\mcc$, and total optical depth response, and the system total 
luminosity light curve. Middle panel: associated light curves. The profile of the disk dissipation parameter $l_{\mathrm{d}}$ is shown in solid line. Energy bands used are given in keV. All curves are normalised to their maximum. Bottom: Time average spectrum. The primary emission is shown in dotted line, the  
 reflection component in dashed line.\label{simu08}}
\end{figure}

the Schwarschild radius of a 10 $M_{\sun}$ black hole, the light crossing time is $R_{\mathrm{g}}/c= 10 ^{-4} s$. We thus need flare durations which are
only one order of magnitude larger than the light crossing time.
 On the other hand, the invoked dissipation mechanism,
   magnetic reconnection, is very fast, due to the high speed of Alfv\`en waves
   which approaches the speed of light. The reconnection time could 
 be of a few $H/c$ only. There are thus indications that
 at least a fraction
  of the luminosity could be emitted under non-equilibrium conditions.

Let us now consider the case of a short flare in the corona. We assume that the system is initially in a steady state where the corona is not 
(or almost not) powered.
 At a given time, a strong dissipation occurs in the corona.
We model this flare by increasing abruptly $l_{\mathrm{c}}$. The influence of the temporal profile $l_{\mathrm{c}}(t)$, the characteristic duration and amplitude of the dissipation are then to be studied.

\subsection{Results of simulations}

As an example, we take the following profile for the temporal dependence of 
the dissipation parameter:
\begin{eqnarray}
l_{\mathrm{c}}(t)&=&0\qquad \mathrm{when} \quad t<2\nonumber\\
l_{\mathrm{c}}(t)&=&A(t-2)^2\exp\left(-\frac{t-2}{t_{\mathrm{c}}}\right) \qquad \mathrm{when} \quad t>2\label{dissip1}
\end{eqnarray}
Fig.~\ref{simu02} shows the coronal parameters response for $A=1$, $t_{\mathrm{c}}=  H/c$.   At the beginning of the shot  the radiative losses are negligible since 
the photon energy density
 is very weak. The energy dissipation leads to a quick increase 
of the coronal temperature.
  
 A higher temperature means higher Compton losses. 
But as the thermal plasma is still
 photon-starved, cooling is not efficient and the temperature still increases.
 However, a  high energy photon component 
 forms in the medium. Approximately half of these photons, directed upward 
escapes. The others travel toward the disk.
 There, they are instantaneously reprocessed and reinjected in the corona
 as soft radiation.
 The radiative energy is conserved through reprocessing.
 However, a few high energy photons are transformed in numerous soft photons 
which are able to cool the plasma efficiently.
 Then the temperature decreases and the escaping photon spectrum softens.
This feedback mechanism is delayed in time of a few $H/c$ due to the photon
 travel time effect. Thus, if the dissipation parameter increases significantly over
 this time scale, the system cannot adjust gradually and a brutal temperature
increase is unavoidable.

The time (and angle) averaged spectrum of the flare is  shown in
  Fig.~\ref{simu02}.
This spectrum is roughly similar to those we get in stationary situations;
 however there are some important differences that we have to point out.

 The maximum temperature
during the flare ($kT_{\mathrm{e, max}}\sim 400$ keV) is about 4 times higher than the
allowed maximum temperature in steady situations for a slab geometry. 
 As a consequence the overall averaged 
spectral shapes differ.
In the steady state situation, for optical depth, $\taut=0.4$, the equilibrium temperature is only 87 keV, leading to a sharp cut-off in the spectrum above 100 keV.
The intrinsic 4-20 keV slope of the angle averaged spectra is then $\Gamma=2.04$.
 The spectrum shown in Fig~\ref{simu02} has a slightly harder spectral slope 
$\Gamma=1.89$.
 It also extends to higher energies than the steady state spectrum does.
In the example of Fig.~\ref{simu02} about 9 \% of the total luminosity
is emitted above 200 keV, 
while in the steady state case this fraction is only of 2 \%.

The associated light curves shown in Fig.~\ref{simu02} exhibit the
hard-to-soft spectral evolution of the flare. We can also note the 
time delay between the light curves and the dissipation profile.
 This delay is due to photon trapping and multiple reflections in
 the disk/corona.

The maximum temperature achieved depends mainly on
 the dissipation amplitude $A$ and characteristic 
time $t_{\mathrm{c}}$. Higher temperatures are obtained for 
higher amplitudes and shorter durations. The maximum 
temperature is also higher when the initial 
photon energy density is small (i.e. small dissipation 
parameter $l_{\mathrm{d}}$). By manipulating these parameters 
one can thus obtain very large temperature jumps. 
For instance, Fig~\ref{simu03} shows the 
evolution of the parameters during a flare similar to that of Fig.~\ref{simu02}
 with the amplitude~$A$, however, larger by a factor of 20.
 The maximum temperature is around 1.5 MeV. At such a
high temperature, numerous photons with energy higher than 
the pair production threshold are produced.
 Pair production then increases the optical depth by a factor
 of two. After the end of the dissipation, it takes a 
few $H/c$ for these pairs to annihilate. This annihilation 
process leads to the formation of an annihilation 
line in the average spectrum of the flare (shown in Fig.~\ref{simu03}).
Note also the high energy tail which extends up to MeV energies. This tail 
is formed at the beginning of the flare when the temperature is very 
 \clearpage high.
However, due to the important increase of the optical depth during the flare, the   4-20 keV spectral slope is softer ($\Gamma=2.07$) than in the example of Fig.~\ref{simu02}.  

 The light curves presented in Fig~\ref{simu03} are very 
similar to those of the previous example (Fig.~\ref{simu02}).
 The lag between hard and soft photons 
is a general feature of the corona where heating 
is too fast to enable a quasi-static evolution.

This jump in temperature always occurs at the beginning
of the dissipation since the soft photon energy density 
 is then minimum.    
After the bulk of energy has been dissipated, it takes $\sim 10-20 H/c$ for
 it to leave the corona as radiation and for the system to relax.

We can try another shape for the dissipation profile. Fig.~\ref{simu09} shows
 the evolution of the coronal parameters as a response to a dissipation with exponential rise and instantaneous decay:
\begin{eqnarray}
l_{\mathrm{c}}(t)&=&0\qquad \mathrm{when} \quad t<2 \qquad{\mathrm{and}} \qquad t>2+t_{\mathrm{c}}\nonumber\\  
l_{\mathrm{c}}(t)&=&A\frac{\mathrm{e}^{(t-2)/t_{\mathrm{c}}}-1}{\mathrm{e}-1} \qquad 
\mathrm{when} \quad 2<t<2+t_{\mathrm{c}}\label{dissip2}
\end{eqnarray}


The shot amplitude and duration have  been fixed respectively at $A=100$ 
and $t_{\mathrm{c}}=10$ $H/c$. The optical depth associated with ions is $\taus=1$.
The temperature, after having reached
 its maximum ($\sim$ 550 keV) at the very beginning of the flare, decreases to its steady 
state equilibrium value ($\sim$ 50 keV) as the dissipation parameter
 still rises.

The initial temperature increase leads to an intense pair production  
raising the optical depth by $\Delta\taut=0.6$. The annihilation of the 
pair excess does not form an annihilation line. Due to the  
high impulse amplitude,
the line emissivity is negligible compared to the continuum high energy flux.
Thus a large increase of the optical depth during the flare does not systematically involve the formation of an annihilation line.

 The flare light curves are shown in Fig.~\ref{simu09}. One can see that
 the 200-2000 keV  light curve reaches its maximum at the very beginning 
of the shot unlike the lower  energy bands which follow the dissipation 
curve. This peak is obviously linked with the temperature maximum. Here again
the spectral evolution leads to soft-photon lags.

If the shot duration is long ($t_{c}\gg 1$), as in the example of 
Fig.~\ref{simu08} where $A=10$ and $t_{c}=100 H/c$, the system is in 
quasi equilibrium state during 80\% of the flare duration. 
The radiation produced during the initial temperature excess is 
negligible in the integrated spectrum. This resulting spectrum
 (shown in Fig.~\ref{simu08}) is the same as the one obtained 
in the steady state approximation. The 4-20 keV slope is $\Gamma=2.33$;
 it is then interesting to compare this spectrum to that of
 Fig.~\ref{simu09} which is harder ($\Gamma=2.00$) and presents a 
high energy tail.

Although it is not a characteristic linked with the dynamics, we can also
 note that the large optical depth of order unity considered here 
wipes out the reflection features, reducing the apparent
 amount of reflection.

\subsection{Discussion}

Our simulations show that when the dissipation time is short,
the spectral evolution during the flare evolves from hard to soft 
producing soft-photon lags. 
As expected, a fast dissipation induces a temperature 
jump at the beginning of the flare, followed by a cooling.

  This arises from the 
response time of the disk due to photon travelling in the corona.
The slab corona geometry considered here has the shortest response time.
 Other geometries where the corona is physically 
separated from the disk 
have a longer feedback time.  We thus expect this  effect
 to be amplified in such configurations.

 Such soft lags
are not compatible with hard lags generally observed in X-ray binaries
 (e.g., Cui 1999 and references 
therein\nocite{1999hepa.conf...97C}).
Observations, however, generally show hard
 lags in X-ray binaries only up to 30 Hz.  The lags
cannot presently be measured at higher frequencies. 

 If we assume that the size of
 the corona is of the order of 
the Schwarschild radius of a 10 $M_{\sun}$ black hole, the light crossing time is $R_{\mathrm{g}}/c= 10 ^{-4} \mathrm{s}$.
The observed lags are thus  associated with variability 
on time scales greater than 50 $R_{\mathrm{g}}/c$, probably large enough for   
a quasi-static evolution to take place.

This low frequency variability could be due to long (quasi-static) 
independent  events, then the observed hard lags would be the 
 consequence of the  individual flare spectral evolution which is then 
quasi-steady. 
More likely, the low frequency variability could be due to a succession of 
correlated shorter events (Poutanen \& Fabian 1999).
\nocite{1999MNRAS.306L..31P}
The observed time lags would then depend on the spectral evolution 
of the avalanche and not necessarily on the individual flare 
spectral evolution. 

  Each short flare event
 can then have a hard to 
soft  spectral evolution.
The only constraint, in order to be consistent with the observations,
is that the flares should be on average softer at the beginning,
and  harder at the end of the avalanche.
   
Anyway models where variability is driven by the dissipation in the corona 
predict that the lags should invert from hard to soft lags at some frequency
 related to the size of the emitting region. We showed that a dissipation
 occuring on time scales of $\sim 10 H/c$ can produce soft lag while 
dissipations on time scales $\sim 100 H/c$ enable a quasi-steady state
 evolution. Thus the lag inversion lies somewhere in between these
 two time scales. If we assume $H/c\sim R_{\mathrm{g}}/c\sim 10^{-4} \mathrm{s}$, then the lag inversion occurs below $\sim 150$ Hz. In other words, the observation
of such a lag inversion would put constraints on the 
size of the emitting region. Detailed Monte-Carlo simulations can
help in determining precisely where the lag inversion should fall, depending mainly on geometry.

We can also note that coronal flare heating and subsequent cooling by reprocessing might explain the
soft lags observed in neutron star systems such as the millisecond pulsar \object{SAX J1808.4-3658} 
(Cui et al., 1998)\nocite{1998ApJ...504L..27C} or in the kHz QPO
of  \object{4U 1608-52} (Vaughan et al., 1997, 1998\nocite{1997ApJ...483L.115V}\nocite{1998ApJ...509L.145V}).

In addition, we showed that the dynamical processes can also have 
an influence on the average spectra. As a consequence of the rise 
in temperature, which reaches higher values than in steady state situations, 
it is possible to obtain harder
spectra. This could relax somewhat the geometry constraints
 on spectral shape.

With convenient parameters, this thermal model can even produce annihilation lines.
The details of the physics of accretion disk coronae are presently unknown. We have thus very few physical constraints, and the parameters required in order to get an annihilation line seem to us plausible. The line appears however for a limited range of parameter values. An intense pair production rate is required at the beginning of the flare and thus a huge dissipation on very short time scales. If, however, the dissipation amplitude is too large the line disappears because of the strong luminosity in the continuum around 511 keV.
   
The fact that these lines 
are  not  observed (see however Bouchet et al. 1991;\nocite{1991ApJ...383L..45B} Goldwurm et al. 1992\nocite{1992ApJ...389L..79G} ) is consistent with a nearly steady situation in compact sources, but does not imply it.       

Another interesting feature is the formation of a high-energy tail at MeV energies. The observed tail in Cyg X-1 (Ling et al. 1997\nocite{1997ApJ...484..375L}; McConnell et al. 1997\nocite{1997comp.symp..829M}) can probably be explained by pure
 non-equilibrium thermal comptonisation without need for hybrid models.

\section{Conclusion}
The NLMC method has significantly contributed  to the study of the
 spectral properties of a steady corona radiatively coupled with an
 accretion disk (e.g. Stern et al. 1995b\nocite{1995ApJ...449L..13S}).
However, in order to understand the short term X-ray variability of 
accreting black hole sources, it seems necessary to take into account the dynamical aspect of the coupling. We have shown that, here again, 
the NLMC method can be an efficient tool.
Our code is able to deal with different situations in which the 
disk-corona equilibrium is perturbed by a violent energy dissipation 
in the disk or the corona.
The few examples given here are far from being a definitive study of these 
problems. However, they enable us to outline some important general 
properties.

On the one hand, we showed that models invoking a variability in 
the injection of soft seed photons as the origin of hard X-ray
 variability have to take into account the response of the corona
 to such fluctuations. Indeed, the corona is quickly Compton cooled when
 the soft photons flux increases. As long as the coronal heating is kept
 constant, the
whole luminosity of the Comptonised radiation is constant, even if the thermal
 emission is strongly variable.
There is however an important spectral evolution, leading to complex correlations
between different energy bands. Details of these correlations depend on 
the intrinsic soft flux variability.

On the other hand, if the variability arises from
dissipation in the corona, the quasi-static approximation is valid 
as long as the dissipation time scale is far larger than the corona
 light crossing
time. If this is not the case the feedback from the disk leads to a hard-to-soft 
spectral evolution. Such models thus predict soft lags at high fourier 
frequencies ($> \sim$ 150 Hz). We also showed that a short dissipation
 time scale produces harder spectra than steady state dissipation.
Moreover, in such conditions, a pure thermal model can
 produce spectral features 
such as high-energy tails or annihilation lines which are generally 
considered as the signature of non-thermal processes.

The present work could be 
developed in numerous ways.
For example, as the disk and corona may be coupled by some physical mechanism, 
it is likely that the observed variability originates from nearly simultaneous 
perturbations of the disk
and the corona. It is also not very realistic to consider that the 
dissipation occurs 
both homogeneously and instantaneously in the corona or the disk: 
effects of propagation should be introduced.
Other complications may arise, such as bulk motions or modifications of the
 geometry of the emitting region during a flare. 
The studies of the individual flare evolution
should provide predictions for the lag inversion frequencies that 
can be used to put constraints on the geometry of the sources.

Then, it would be interesting to investigate different stochastic
 models  describing  the  interaction   between  flares.
  Taken  together  with  the  evolution  of
  individual  flares it would make it possible to generate  the light curves
  in  different  energy  bands, and compute the  time-averaged  energy
  spectra  and  various  temporal  characteristics  such as power  spectral
  density,    time/phase    lags   between    different    energy    bands,
  cross-correlation  functions,  coherence function, and compare them
  with the observations.

\begin{acknowledgements}
 We are grateful to Boris Stern for providing us with the results of his code
 for comparisons. We thank Juri Poutanen for a critical reading of the 
manuscript and many useful comments.
\end{acknowledgements}


\begin{thebibliography}{}

\bibitem[\protect\astroncite{{Barbosa}}{1982}]{1982ApJ...254..301B}
{Barbosa} D.~D., 1982,
\newblock {\apj} {254}, 301

\bibitem[\protect\astroncite{{Beloborodov}}{1999}]{1999ApJ...510L.123B}
{Beloborodov} A.~M., 1999,
\newblock {\apjl} {510}, L123

\bibitem[\protect\astroncite{{Böttcher} and
  {Liang}}{1998}]{1998ApJ...506..281B}
{Böttcher} M., {Liang} E.~P., 1998,
\newblock {\apj} {506}, 281

\bibitem[\protect\astroncite{{Böttcher} and
  {Liang}}{1999}]{1999ApJ...511L..37B}
{Böttcher} M., {Liang} E.~P., 1999,
\newblock {\apjl} {511}, L37

\bibitem[\protect\astroncite{{Bouchet} et~al.}{1991}]{1991ApJ...383L..45B}
{Bouchet} L., {Mandrou} P., {Roques} J. P. et al., 1991,
\newblock {\apjl} {383}, L45


\bibitem[\protect\astroncite{{Coppi}}{1992}]{1992MNRAS.258..657C}
{Coppi} P.~S., 1992,
\newblock {\mnras} {258}, 657

\bibitem[\protect\astroncite{{Coppi} and
  {Blandford}}{1990}]{1990MNRAS.245..453C}
{Coppi} P.~S., {Blandford} R.~D., 1990,
\newblock {\mnras} {245}, 453

\bibitem[\protect\astroncite{{Cui}}{1999}]{1999hepa.conf...97C}
{Cui} W., 1999,
\newblock in {High Energy Processes in Accreting Black Holes, ASP Conference
  Series 161, ed. J. Poutanen \& R. Svensson.}, p.~97

\bibitem[\protect\astroncite{{Cui} et~al.}{1998}]{1998ApJ...504L..27C}
{Cui} W., {Morgan} E.~H., {Titarchuk} L.~G., 1998,
\newblock {\apjl} {504}, L27

\bibitem[\protect\astroncite{{Done} and {Fabian}}{1989}]{1989MNRAS.240...81D}
{Done} C., {Fabian} A.~C., 1989,
\newblock {\mnras} {240}, 81

\bibitem[\protect\astroncite{{Dove} et~al.}{1997a}]{1997ApJ...487..747D}
{Dove} J.~B., {Wilms} J., {Begelman} M.~C., 1997a,
\newblock {\apj} {487}, 747

\bibitem[\protect\astroncite{{Dove} et~al.}{1997b}]{1997ApJ...487..759D}
{Dove} J.~B., {Wilms} J., {Maisack} M., {Begelman} M.~C., 1997b,
\newblock {\apj} {487}, 759

\bibitem[\protect\astroncite{{Edelson} and
  {Nandra}}{1999}]{1999ApJ...514..682E}
{Edelson} R., {Nandra} K., 1999,
\newblock {\apj} {514}, 682

\bibitem[\protect\astroncite{{Fabian} et~al.}{1986}]{1986MNRAS.221..931F}
{Fabian} A.~C., {Guilbert} P.~W., {Blandford} R.~D., {Phinney} E.~S.,
  {Cuellar} L., 1986,
\newblock {\mnras} {221}, 931

\bibitem[\protect\astroncite{{Galeev} et~al.}{1979}]{1979ApJ...229..318G}
{Galeev} A.~A., {Rosner} R., {Vaiana} G.~S., 1979,
\newblock {\apj} {229}, 318

\bibitem[\protect\astroncite{{George} and {Fabian}}{1991}]{1991MNRAS.249..352G}
{George} I.~M., {Fabian} A.~C., 1991,
\newblock {\mnras} {249}, 352

\bibitem[\protect\astroncite{{Ghisellini} et~al.}{1993}]{1993MNRAS.263L...9G}
{Ghisellini} G., {Haardt} F., {Fabian} A.~C., 1993,
\newblock {\mnras} {263}, L9

\bibitem[\protect\astroncite{{Goldwurm} et~al.}{1992}]{1992ApJ...389L..79G}
{Goldwurm} A., {Ballet} J., {Cordier} B. et al., 1992,
\newblock {\apjl} {389}, L79

\bibitem[\protect\astroncite{{Gorecki} and
  {Wilczewski}}{1984}]{1984AcA....34..141G}
{Gorecki} A., {Wilczewski} W., 1984,
\newblock {Acta Astronomica} {34}, 141


\bibitem[\protect\astroncite{{Guilbert} and {Rees}}{1988}]{1988MNRAS.233..475G}
{Guilbert} P.~W., {Rees} M.~J., 1988,
\newblock {\mnras} {233}, 475

\bibitem[\protect\astroncite{{Guilbert} and
  {Stepney}}{1985}]{1985MNRAS.212..523G}
{Guilbert} P.~W., {Stepney} S., 1985,
\newblock {\mnras} {212}, 523

\bibitem[\protect\astroncite{{Guilbert} et~al.}{1983}]{1983MNRAS.205..593G}
{Guilbert} P.~W., {Fabian} A.~C., {Rees} M.~J., 1983,
\newblock {\mnras} {205}, 593

\bibitem[\protect\astroncite{{Haardt} and
  {Maraschi}}{1991}]{1991ApJ...380L..51H}
{Haardt} F., {Maraschi} L., 1991,
\newblock {\apjl} {380}, L51

\bibitem[\protect\astroncite{{Haardt} and
  {Maraschi}}{1993}]{1993ApJ...413..507H}
{Haardt} F., {Maraschi} L., 1993,
\newblock {\apj} {413}, 507

\bibitem[\protect\astroncite{{Haardt} et~al.}{1994}]{1994ApJ...432L..95H}
{Haardt} F., {Maraschi} L., {Ghisellini} G., 1994,
\newblock {\apjl} {432}, L95

\bibitem[\protect\astroncite{{Haardt} et~al.}{1997}]{1997ApJ...476..620H}
{Haardt} F., {Maraschi} L., {Ghisellini} G., 1997,
\newblock {\apj} {476}, 620

\bibitem[\protect\astroncite{{Hua} et~al.}{1997}]{1997ApJ...482L..57H}
{Hua} X.~M., {Kazanas} D., {Titarchuk} L., 1997,
\newblock {\apjl} {482}, L57

\bibitem[\protect\astroncite{{Ichimaru}}{1977}]{1977ApJ...214..840I}
{Ichimaru} S., 1977,
\newblock {\apj} {214}, 840

\bibitem[\protect\astroncite{{Jourdain} and
  {Roques}}{1995}]{1995ApJ...440..128J}
{Jourdain} E., {Roques} J.~P., 1995,
\newblock {\apj} {440}, 128

\bibitem[\protect\astroncite{{Kazanas} et~al.}{1997}]{1997ApJ...480..735K}
{Kazanas} D., {Hua} X.~M., {Titarchuk} L., 1997,
\newblock {\apj} {480}, 735

\bibitem[\protect\astroncite{{Kusunose}}{1987}]{1987ApJ...321..186K}
{Kusunose} M., 1987,
\newblock {\apj} {321}, 186

\bibitem[\protect\astroncite{{Li} et~al.}{1996}]{1996A&AS..120C.167L}
{Li} H., {Kusunose} M., {Liang} E.~P., 1996,
\newblock {\aaps} {120}, C167

\bibitem[\protect\astroncite{{Liang} and {Price}}{1977}]{1977ApJ...218..247L}
{Liang} E. P.~T., {Price} R.~H., 1977,
\newblock {\apj} {218}, 247

\bibitem[\protect\astroncite{{Ling} et~al.}{1997}]{1997ApJ...484..375L}
{Ling} J.~C., {Wheaton} W.~A., {Wallyn} P. et al.,
  1997,
\newblock {\apj} {484}, 375

\bibitem[\protect\astroncite{{Malzac} et~al.}{1998}]{1998A&A...336..807M}
{Malzac} J., {Jourdain} E., {Petrucci} P.~O., {Henri} G., 1998,
\newblock {\aap} {336}, 807

\bibitem[\protect\astroncite{{McConnell} et~al.}{1997}]{1997comp.symp..829M}
{McConnell} M., {Bennett} K., {Bloemen} H. et al., 1997,
\newblock in {AIP Conf. Proc. 410: Proceedings of the Fourth Compton
  Symposium}, p. 829

\bibitem[\protect\astroncite{{Miyamoto} and
  {Kitamoto}}{1989}]{1989Natur.342..773M}
{Miyamoto} S., {Kitamoto} S., 1989,
\newblock {\nat} {342}, 773

\bibitem[\protect\astroncite{{Miyamoto} et~al.}{1988}]{1988Natur.336..450M}
{Miyamoto} S., {Kitamoto} S., {Mitsuda} K., {Dotani} T., 1988,
\newblock {\nat} {336}, 450

\bibitem[\protect\astroncite{{Nandra} et~al.}{1998}]{1998ApJ...505..594N}
{Nandra} K., {Clavel} J., {Edelson} R.~A. et al., 1998,
\newblock {\apj} {505}, 594

\bibitem[\protect\astroncite{{Narayan} and {Yi}}{1994}]{1994ApJ...428L..13N}
{Narayan} R., {Yi} I., 1994,
\newblock {\apjl} {428}, L13

\bibitem[\protect\astroncite{{Nayakshin} and {Dove}}{1999}]{nayak99}
{Nayakshin} S., {Dove} J., 1999,
\newblock {in press, astro-ph/9811059}

\bibitem[\protect\astroncite{{Nowak} et~al.}{1999a}]{1999ApJ...510..874N}
{Nowak} M.~A., {Vaughan} B.~A., {Wilms} J., {Dove} J.~B.,
  {Begelman} M.~C., 1999a,
\newblock {\apj} {510}, 874

\bibitem[\protect\astroncite{{Nowak} et~al.}{1999b}]{1999ApJ...515..726N}
{Nowak} M.~A., {Wilms} J., {Vaughan} B.~A., {Dove} J.~B.,
  {Begelman} M.~C., 1999b,
\newblock {\apj} {515}, 726

\bibitem[\protect\astroncite{{Payne}}{1980}]{1980ApJ...237..951P}
{Payne} D.~G., 1980,
\newblock {\apj} {237}, 951

\bibitem[\protect\astroncite{{Poutanen} and {Coppi}}{1998}]{1998Poutetcop}
{Poutanen} J., {Coppi} P.~S., 1998,
\newblock {Physica Scripta} {T77}, 57,
\newblock (astro-ph/9711316)

\bibitem[\protect\astroncite{{Poutanen} and
  {Fabian}}{1999}]{1999MNRAS.306L..31P}
{Poutanen} J., {Fabian} A.~C., 1999,
\newblock {\mnras} {306}, L31

\bibitem[\protect\astroncite{{Poutanen} and
  {Svensson}}{1996}]{1996ApJ...470..249P}
{Poutanen} J., {Svensson} R., 1996,
\newblock {\apj} {470}, 249

\bibitem[\protect\astroncite{{Poutanen} et~al.}{1997}]{1997MNRAS.292L..21P}
{Poutanen} J., {Krolik} J.~H., {Ryde} F., 1997,
\newblock {\mnras} {292}, L21

\bibitem[\protect\astroncite{{Pozdnyakov} et~al.}{1983}]{1983SSRvE...2..189P}
{Pozdnyakov} L.~A., {Sobol} I.~M., {Sunyaev} R.~A., 1983,
\newblock {Soviet Scientific Reviews Section E Astrophysics and Space Physics
  Reviews} {2}, 189

\bibitem[\protect\astroncite{{Ross} et~al.}{1999}]{1999MNRAS.306..461R}
{Ross} R.~R., {Fabian} A.~C., {Young} A.~J., 1999,
\newblock {\mnras} {306}, 461

\bibitem[\protect\astroncite{{Shapiro} et~al.}{1976}]{1976ApJ...204..187S}
{Shapiro} S.~L., {Lightman} A.~P., {Eardley} D.~M., 1976,
\newblock {\apj} {204}, 187

\bibitem[\protect\astroncite{{Stern} et~al.}{1995a}]{1995MNRAS.272..291S}
{Stern} B.~E., {Begelman} M.~C., {Sikora} M., {Svensson} R., 1995a,
\newblock {\mnras} {272}, 291

\bibitem[\protect\astroncite{{Stern} et~al.}{1995b}]{1995ApJ...449L..13S}
{Stern} B.~E., {Poutanen} J., {Svensson} R., {Sikora} M., {Begelman}
  M.~C., 1995b,
\newblock {\apjl} {449}, L13

\bibitem[\protect\astroncite{{Sunyaev} and
  {Titarchuk}}{1980}]{1980A&A....86..121S}
{Sunyaev} R.~A., {Titarchuk} L.~G., 1980,
\newblock {\aap} {86}, 121

\bibitem[\protect\astroncite{{Svensson}}{1982}]{1982ApJ...258..321S}
{Svensson} R., 1982,
\newblock {\apj} {258}, 321

\bibitem[\protect\astroncite{{Svensson} et~al.}{1996}]{1996A&AS..120C.587S}
{Svensson} R., {Larsson} S., {Poutanen} J., 1996,
\newblock {\aaps} {120}, C587

\bibitem[\protect\astroncite{{Vaughan} et~al.}{1997}]{1997ApJ...483L.115V}
{Vaughan} B.~A., {van der Klis} M., {M\'endez} M. et al., 1997,
\newblock {\apjl} {483}, L115

\bibitem[\protect\astroncite{{Vaughan} et~al.}{1998}]{1998ApJ...509L.145V}
{Vaughan} B.~A., {Van Der Klis} M., {Méndez} M. et al., 1998,
\newblock {\apjl} {509}, L145

\bibitem[\protect\astroncite{{Zdziarski} et~al.}{1996}]{1996A&AS..120C.553Z}
{Zdziarski} A.~A., {Gierlinski} M., {Gondek} D., {Magdziarz} P., 1996,
\newblock {\aaps} {120}, C553

\end{thebibliography}
\end{document}